\documentclass[p5]{elsarticle}
\usepackage{color,graphicx,threeparttable,amsmath,textcomp,hyperref,selectp}
\usepackage{multirow}
\usepackage[dvipsnames]{xcolor}
\usepackage[official]{eurosym}
\AtBeginDocument{\colorlet{defaultcolor}{.}}

\newlength{\picwidth}
\newcommand{\iotname}{EIoT }

\journal{Internet of Things}

\begin{document}
\begin{frontmatter}

\title{How does enterprise IoT traffic evolve? Real-world evidence from a Finnish operator\tnoteref{conf}}
\tnotetext[conf]{A shorter version of this work was presented at the 2019 IEEE 5th World Forum on Internet of Things and was published in its proceedings \cite{finley2019}.}

\cortext[cor1]{Corresponding author}
\author[helsinki]{B. Finley\corref{cor1}}
\ead{benjamin.finley@helsinki.fi}
\author[aalto]{J. Benseny}
\ead{jaume.benseny@aalto.fi}
\author[mpkk]{A. Vesselkov}
\ead{vesselkov.as@gmail.com}
\author[aalto]{J. Walia}
\ead{jaspreet.walia@aalto.fi}

\address[helsinki]{Department of Computer Science, University of Helsinki, Helsinki, Finland}
\address[aalto]{Department of Communications and Networking, Aalto University, Helsinki, Finland}
\address[mpkk]{Department of Military Technology, National Defence University, Helsinki Finland}



\begin{abstract}
The adoption of Internet of Things (IoT) technologies in businesses is increasing and thus enterprise IoT (EIoT) is seemingly shifting from hype to reality. However, the actual use of \iotname over significant timescales has not been empirically analyzed.  In other words, the reality remains unexplored. Furthermore, despite the variety of \iotname verticals, the use of IoT across vertical industries has not been compared. This paper uses a two-year \iotname dataset from a major Finnish mobile network operator to investigate device use across industries, cellular traffic patterns, and mobility patterns. We present a variety of novel findings: \iotname traffic volume per device has increased three-fold over the last two years, the share of LTE-enabled devices has remained low at around 2\% and that 30\% of \iotname devices are still 2G only, and there are order of magnitude differences between different industries' \iotname traffic and mobility. We also show that daily traffic can be clustered into only three patterns, differing mainly in the presence and timing of a peak hour. Beyond these descriptive results, modeling and forecasting is conducted for both traffic and mobility. {\color{black} We forecast the total daily \iotname traffic through a temporal regression model and achieve an error of about 15\% over medium-term (30 to 180 day) horizons. We also model device mobility through a Markov mixture model and quantify the upper bound of predictability for device mobility.}
\end{abstract}


\end{frontmatter}

\section{Introduction}\label{sec:introduction}
According to a recent survey \cite{AnalysysMason2017}, 29\% of companies globally utilize Internet of Things (IoT) devices, suggesting that IoT use by companies (also referred to as enterprise IoT or EIOT) is shifting from hype to reality. Furthermore, \iotname is gaining momentum across different industries that use IoT devices in unique ways for solving diverse problems. Despite the variety of \iotname verticals and applications, the actual usage of IoT across industries has never been empirically explored. Additionally, the few studies \cite{Marjamaa2012, shafiq2013, Romirer-Maierhofer2015} that have analyzed \iotname device usage from commercial cellular networks are relatively old and have only analyzed short timescales (i.e., typically less than a few weeks). Thus the evolution of \iotname usage over longer timescales remains uninvestigated.

To address these gaps, this work aims to study the evolution of traffic, mobility, and population characteristics of \iotname devices on several different timescales and on both the general and industry-level. Additionally, the work aims to illustrate the feasibility of modeling and predicting several of these diverse characteristics. Impact-wise, traffic, mobility, and population characteristics all are relevant factors for network operators in terms of both technical network operations, network planning, and longer term network investment strategies.

Towards these aims, we analyze a two-year \iotname dataset from a major Finnish mobile network operator (MNO) that includes data traffic volumes, customer industry class, and device features. We first perform a descriptive analysis of the traffic, mobility, and population characteristics while highlighting insights along the way. We then, in terms of modeling and prediction, cluster daily temporal traffic patterns, forecast longer term traffic volumes, model mobility through a Markov mixture model, and calculate the upper bound of mobility predictability (for an ideal model).

Overall, the work gives a holistic view of the evolution and current state of \iotname usage in a major MNO, thus illustrating the reality instead of the hype. We note that Finland is an early \iotname adopter with the 6th most M2M modules per capita of OECD countries (23 per 100 inhabitants) \cite{oecdm2m2017}. {\color{black} We also note that we do not claim the results can be greatly generalized to other operators or countries. Instead we hope that the results can represent a case in a larger process of cross-study comparison between different operators and countries, thus building up general patterns and theory.}

The results of this study are relevant to both researchers and practitioners. In particular, researchers studying the impact of \iotname on cellular networks can use the identified \iotname traffic and mobility patterns to improve modeling. Furthermore, providers of \iotname connectivity and other services can get a better understanding of the requirements and challenges of IoT devices in different verticals, which will allow them to address customer needs. {\color{black}Also we note that the broad and diverse characteristics and methods of the study were chosen to give, as mentioned, a holistic guide such that future work can focus in on more specific details and with more specialized methods.}

We briefly describe the structure of the remainder of the paper. Section \ref{sec:background} summarizes related work on empirical IoT traffic analysis, Section \ref{sec:dataset} describes the dataset, and Section \ref{sec:desc_results} details the basic descriptive results including temporal, spatial (i.e., mobility), and \iotname device population aspects. Section \ref{sec:adv_results} presents further analysis and modeling results covering temporal clustering, temporal forecasting, and mobility modeling. Finally Section \ref{sec:limitations} discusses the limitations and Section \ref{sec:conclusions} the conclusions and implications.

\section{Related Work}\label{sec:background}
Shafiq et al. \cite{shafiq2013} were the first to analyze \iotname\footnote{They denoted such traffic as machine-to-machine (M2M).} data from a commercial cellular network in the US. They examined the traffic generated by more than a million \iotname devices over one week in August 2010 and found that such devices are less mobile than smartphones, generate more uplink than downlink traffic, and often have synchronized activity. Ref. \cite{Romirer-Maierhofer2015} confirmed these last two observations by analyzing \iotname device data collected over several weeks in 2013 by a European mobile operator. Both studies concluded that the traffic generated by \iotname devices significantly differs from smartphones, indicating the need for MNOs to reassess network planning traditionally optimized for smartphone users.

In a more recent study, Andrade et al. \cite{Andrade2017} analyzed the traffic and mobility patterns of one million connected cars on a cellular network in the US. The authors concluded that the data traffic that cars generate differ both from smartphones and other IoT devices, and warned about the potential adverse impact that massive over-the-air firmware updates may have on network performance.

Several studies \cite{Baer2016, Baer2015, Laner2014} similarly analyzed IoT data from a cellular network but with different objectives. The studies proposed methods for online and offline classification of IoT traffic that would give MNOs a more efficient way of identifying IoT devices compared to the traditional TAC-based (Type Allocation Code) approach.

{\color{black}
From a mobility modeling perspective, smartphone mobility has been extensively modeled using empirical data. For example, based on a 13-month dataset, \cite{kim2006} modeled movement between highly visited locations in which transition speeds and pause times follow log-normal distributions. In another example, \cite{lee2006} modeled device location transitions via a semi-Markov process using a transition matrix and transition time distribution. However, as far as we know no study has modeled mobility for general \iotname devices. Though, empirical mobility models for specific IoT devices types (such as vehicles \cite{pigne2011}) have been developed.

Several studies have also modeled longer term (\textgreater1 week) internet traffic volume trends (though only at the aggregate level without differentiating IoT devices). For example, \cite{korotky2013} accurately fit a hyperbolic function to the CAGRs\footnote{Compound annual growth rates} of 20 years of fixed and mobile internet traffic volumes. Ref. \cite{vlachos2016} provides further background into such long term traffic modeling including varying methods, timescales, and datasets.}

\section{Dataset}\label{sec:dataset}
{\color{black} Before describing the dataset, we first note that an \iotname device (e.g., smart-meter, asset tracker, etc.) typically contains a generic communication module (CM) to transmit the data the device collects. Since these modules are often integrated, a device naturally inherits properties of the CM such as network capabilities. Therefore, in this work differentiating between properties of the \iotname device and CM is only required in a few cases. Hereafter, we note specifically when this is the case.}

The main dataset of the analysis is a collection of data detail records (DDRs) of devices that use an IoT-specific enterprise subscription provided by a major nation-wide Finnish MNO. In other words, in this work, an \iotname device is defined as a device that uses a IoT-specific enterprise subscription (subscription based definition).

The dataset covers a period of two years from September 2016 to August 2018. Each record covers a single hour and contains the following fields: anonymized IMSI\footnote{We only refer to devices in this work and we assume a one-to-one relationship between IMSI and device since SIM cards are rarely swapped to different devices. Empirically we find that only 1.6\% of IMSIs were used with multiple devices over the entire period.}, anonymized cell ID, anonymized customer ID (hereafter company ID), device TAC, uplink traffic volume (in bytes), and downlink traffic volume (in bytes). If the device had traffic in more than one cell in a given hour then additional records for that hour for each cell were included. In other words, each record is uniquely identified by a triplet of (device, cell, hour). The dataset was extracted from the operator's network accounting system which receives aggregate statistics from base stations. We also note that the hourly time granularity of the dataset is a result of collecting the dataset from this network accounting system.

Additionally, the DDR dataset was joined with two other MNO provided datasets: a dataset of device features (from the GSMA device database) for all TACs found in the DDR dataset and a dataset of company industries for each company ID in the DDR dataset. The device feature dataset includes the following fields: device CM model name, device CM release year, and device network capabilities (i.e., EDGE, HSPA, LTE, etc.), while the company industry dataset is based on the standard Finnish TOL2008\footnote{TOL2008 is based on the EU's classification of economic activities, NACE Rev.2 \cite{Eurostat2008}, prescribed in EC Regulation No 1893/2006} industry classification. For industry-level analyses, we only include industries with at least 10 distinct companies and where the largest company accounts for a maximum of 80\% of traffic or devices in the industry. For reference, we list these industries, their acronyms (used in figures), and brief descriptions in Table \ref{tab:industry_descriptions}. 

As previously mentioned, in this work, an \iotname device is defined as a device that uses a enterprise IoT-specific subscription (i.e., subscription based definition). However, to further ensure that only \iotname devices were included in the analysis, we first manually checked all unique device CM models from the dataset and categorized them as IoT, maybe-IoT (typically PCI Express data cards that can also be used in laptops), and non-IoT (typically smartphones and feature phones) based on online research. We then filtered out all non-IoT devices and any device with an invalid TAC code (since in those cases we did not have any device information). This filtering removed 5.7\% of devices.

We also note that since we only know the CM model name (and not the \iotname device model name), we do not know the \iotname device type. For example, we might know that a device has a Cinterion EU3-E module, but we do not know if the device is a smart meter, asset tracker, etc. Unfortunately inferring the device type from just the CM is not feasible because, as mentioned, these modules are generic and many manufacturers do not identify the CMs in their devices. We even attempted to scrape FCC and other regulatory approval reports to identify the CMs in devices but with limited success.

In any case, we know that the device population includes EIoT devices like payment terminals, smart-meters, location trackers, and surveillance cameras. In terms of requirements, some of these devices require very little bandwidth. Payment terminals, for example, typically use the ISO-8583 financial message standard which requires only several kilobytes per payment transaction or even less \cite{iso2003}. While other devices such as video surveillance cameras can require over one Mbps depending on the resolution (e.g., a 720p 30 frame per second H.264 camera requires about 1.9 Mbps). Overall, the analysis further illustrates this requirement diversity, especially across industries.

Finally, to give an idea of the full scale of the analysis, the DDR dataset covers hundreds of companies, hundreds of thousands of devices, and tens of millions of records. We also note that for business confidentiality and privacy reasons we normalize some of the numerical results, however, this normalization does not change the interpretations or conclusions. Finally, for illustration purposes we use moving averages\footnote{The moving average is essentially a low-pass filter in signal processing.} in several figures to help emphasize longer-term trends and smooth out short-term fluctuations.

\begin{table*}[ht]
\caption{Description of industries based on NACE Rev.2 \cite{Eurostat2008}}
\label{tab:industry_descriptions}
\begin{tabular}{p{3.8cm}cp{8.2cm}}
\hline
\textbf{Industry (abbreviated)} & \textbf{Acronym} & \textbf{Description} \\ \hline
Administrative and support & AS & Activities supporting general business operations, except professional activities; e.g., rental and leasing, recruitment, security and investigation \\\hline
Electricity and gas & EG & Providing electric power, natural gas, steam, hot water and the like through a permanent infrastructure \\\hline
Information and communication & IC & Publishing activities, including SW; broadcasting; telecommunications and IT activities \\\hline
Manufacturing & MF & Physical or chemical transformation of materials or components into new products, e.g., food, textiles, computers and electronics \\\hline
Professional activities & PA & Activities making specialized knowledge available to users, e.g., consultancy and engineering \\\hline
Transportation & TR & Provision of passenger or freight transport, and associated activities such as terminal and parking facilities, cargo handling, and storage \\\hline
Wholesale and retail trade & WR & Wholesale and retail sale of any goods, including associated operations, such as assembling and packing; repair of motor vehicles and motorcycles \\\hline
\end{tabular}%

\end{table*}
\section{Descriptive Results}\label{sec:desc_results}
\subsection{Traffic statistics}
\begin{figure}
\centering
\includegraphics[width=\picwidth]{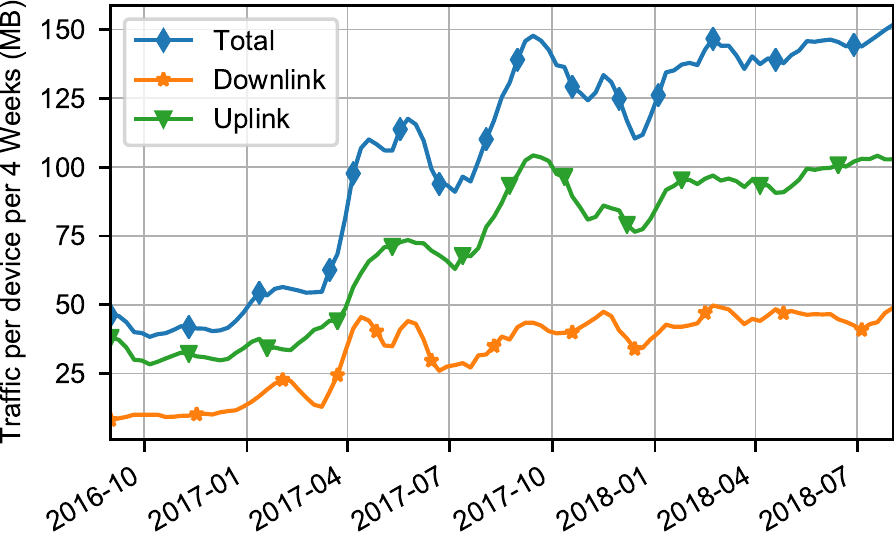}
\caption{Traffic per device per four week period}
\label{fig:data_per_imsi_two_years}
\end{figure}

First, we examine the traffic of cellular \iotname devices over time to understand its evolution. Figure \ref{fig:data_per_imsi_two_years} shows the four-week moving average of traffic per device. We find that total \iotname traffic per device increased three-fold, whereas downlink traffic increased six-fold. Most of the traffic growth occurred between September 2016 and 2017. Comparatively, \cite{shafiq2013} reported a total \iotname traffic increase of 250\% during 2011. Furthermore, despite some fluctuations, the traffic does not demonstrate any seasonal patterns.

\begin{figure}
\centering
\includegraphics[width=\picwidth]{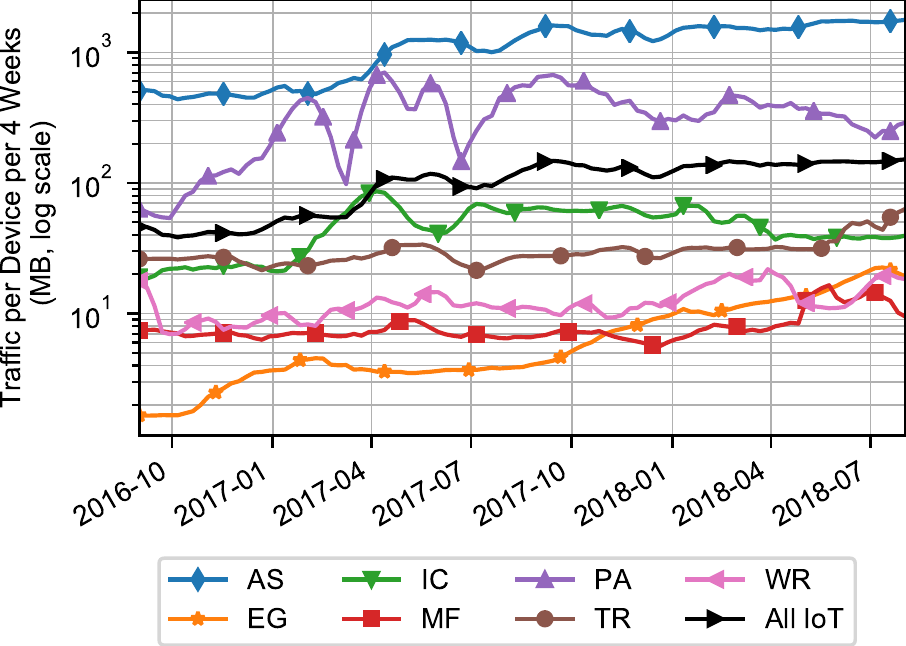}
\caption{Traffic per device per four week period for industries}
\label{fig:data_per_imsi_industry_two_years}
\end{figure}

The volume of traffic increased in all industries, as shown by the four-week moving average of traffic per device in Figure \ref{fig:data_per_imsi_industry_two_years}. We observe significant differences in traffic volumes between industries, with devices in {\it Manufacturing} and {\it Administrative and support} generating on average 10 MB and 2 GB per device, respectively. This difference might be explained by {\it Administrative and support} containing security companies that may generate video traffic via surveillance cameras. The most substantial increase occurred in {\it Electricity and gas} where traffic grew twelve-fold to 19 MB per device. This could be due to a 1-hour metering obligation deadline for smart meters in Finland. We also analyzed the traffic evolution for the subset of companies that had active IoT devices in both the first and last months of the observation period and found similar trends. This indicates that the increase in traffic over time includes both companies that already use IoT and new companies adopting IoT.

\begin{figure}
\centering
\includegraphics[width=\picwidth]{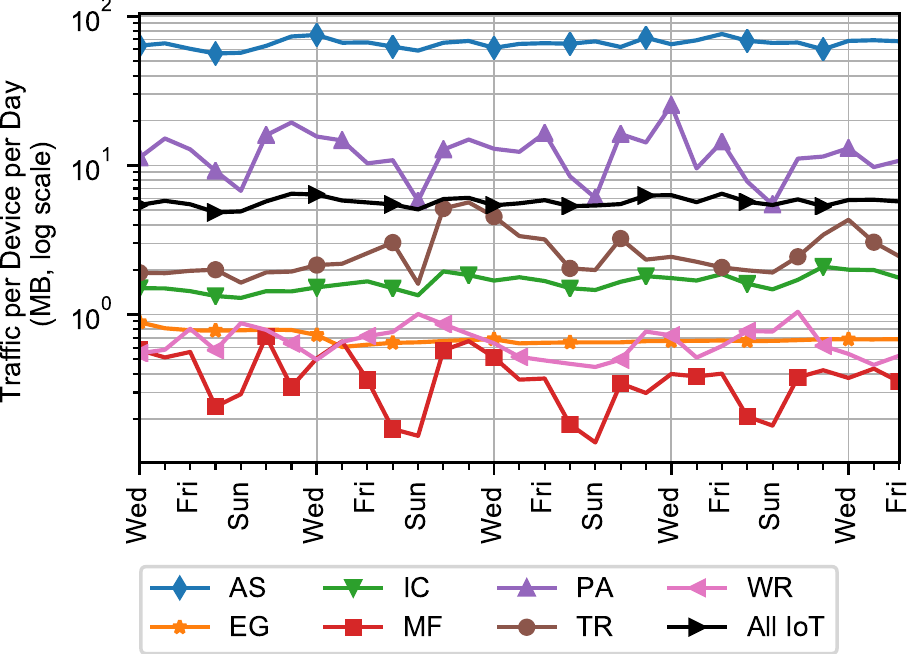}
\caption{Traffic per device per day for August 2018}
\label{fig:data_per_day}
\end{figure}

To explore industry-specific differences in traffic depending on the day of the week, we study the daily traffic per device for August 2018. Figure \ref{fig:data_per_day} illustrates this traffic. We find that most industries do not show significant variation depending on the day of the week. However, in {\it Professional activities} and {\it Manufacturing} industries, we observe weekday-weekend patterns, with traffic halving during weekends.  

\begin{figure}
\centering
\includegraphics[width=\picwidth]{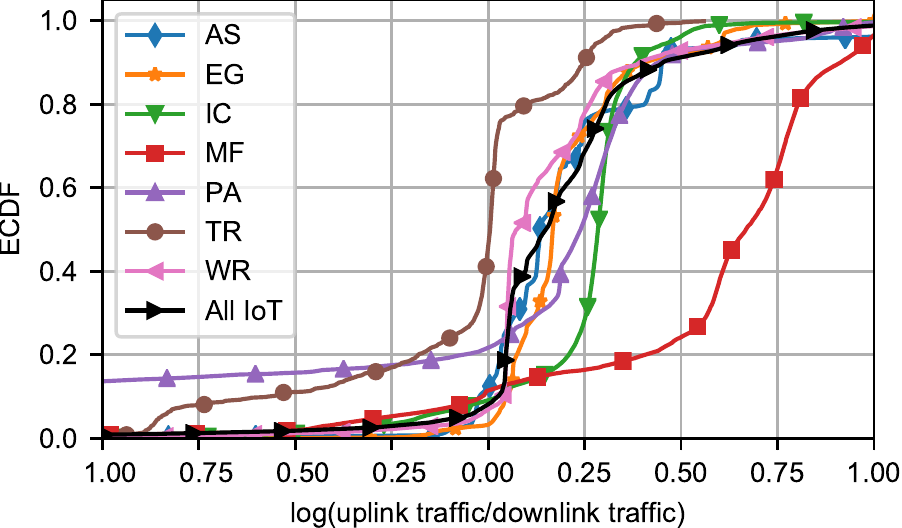}
\caption{ECDF of the log of uplink to downlink traffic ratio for August 2018}
\label{fig:Z_value}
\end{figure}
Furthermore, we study uplink vs. downlink traffic volumes across industries. Figure \ref{fig:Z_value} illustrates the empirical cumulative distribution (ECDF) of the log of uplink/downlink ratio for August 2018. Negative values indicate larger downlink traffic than uplink and positive values vice versa. As the figure shows, 92\% of \iotname devices generated more uplink than downlink traffic, which is consistent with the finding of \cite{shafiq2013}, but exceeds the observation of \cite{Romirer-Maierhofer2015} by more than 30\%. However, in some industries, particularly {\it Transportation} and {\it Professional activities}, the share of devices with greater uplink than downlink traffic is lower, around 54\% and 78\% respectively. Further, in {\it Manufacturing}, the uplink traffic is much larger than downlink traffic (compared to other industries), with a median ratio of 4.66 compared to 1.41 for all \iotname devices. Overall, the results illustrate both intra and inter-industry variation that helps illustrate the diversity of \iotname.

\subsection{Mobility statistics}
\begin{figure}
\centering
\includegraphics[width=\picwidth]{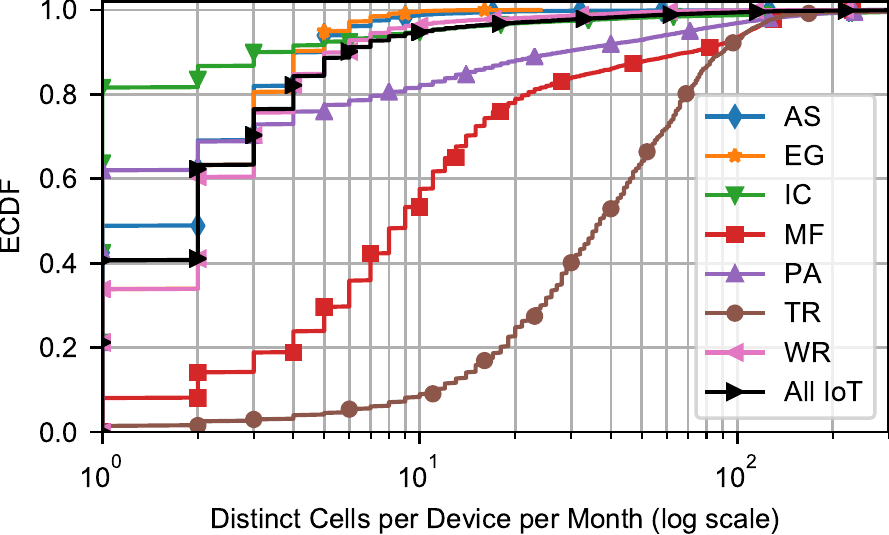}
\caption{ECDF of \iotname device mobility in August 2018}
\label{fig:mobility_CDF}
\end{figure}

Concerning device mobility, we infer such mobility through the number of different cells visited\footnote{The definition of visit only includes cells where traffic was sent or received and thus it is a lower bound on the number of cells attached to by the device. \label{def_ft}} by devices. Figure \ref{fig:mobility_CDF} presents the ECDF of the number of unique cells visited by devices in August 2018. As the figure shows, about 40\% of \iotname devices visited only a single cell indicating significant immobility. Furthermore, for a one week time frame (last week of August), we find an even higher fraction of 55\% of devices visited only a single cell. In comparison, \cite{shafiq2013} found that 30\% of devices visited only a single cell in their one-week dataset. The actual share of stationary devices (again given our definition from footnote \ref{def_ft}) may be even higher since some devices at cell edges may jump between cells depending on signal strength fluctuations or cell load balancing.

We also observe differences in device mobility across industries. Overall around 95\% of all \iotname devices and most devices in {\it Electricity and gas}, {\it Wholesale and retail trade}, and {\it Administrative and support} industries visited less than ten cells per month. Contrastingly and expectedly, devices in the {\it Transportation} industry are very mobile, with 90\% having visited more than ten cells per month. Some industries, for example {\it Manufacturing}, include a mix of mobile and stationary devices. 

\subsection{Cell statistics}
\begin{figure}
\centering
\includegraphics[width=\picwidth]{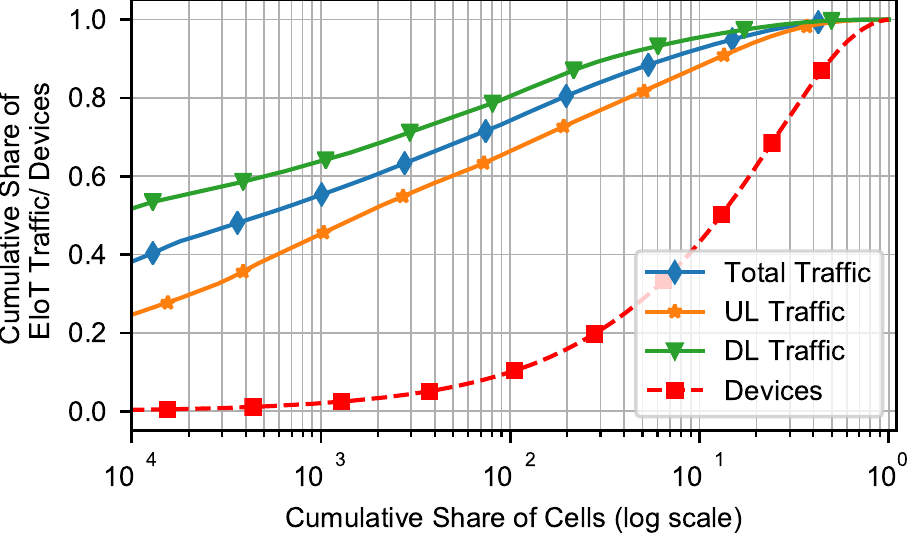}
\caption{Distribution of traffic and devices among cells for August 2018}
\label{fig:spatial_analysis}
\end{figure}

We analyze the distribution of devices and traffic over all the visited cells. Figure \ref{fig:spatial_analysis} illustrates the ECDF of \iotname traffic and devices\footnote{We note that devices are only counted in their most visited cell (in terms of hours), though other definitions such as counting devices in all their visited cells produce similar results.} across \iotname-visited cells in August 2018. The traffic is highly concentrated spatially, with 10\% of cells carrying about 93\% of total \iotname traffic. Comparatively, \cite{paul2011} found 10\% of cells carrying about 55\% of total network traffic in a nationwide network in 2007. The high concentration of \iotname devices and traffic can be explained, given the typical centralization of company campuses compared to normal consumers. This concentration is important for network planning because the deployment of \iotname-specific network features or optimizations would require changes to far fewer cells (and thus cost less) than for non-\iotname features. Finally, in terms of devices, we find that 10\% of cells account for about 44\% of devices while 50\% of cells about 90\% of devices, showing only moderate spatial concentration.

\subsection{\iotname device population statistics}
\begin{figure}
\centering
\includegraphics[width=\picwidth]{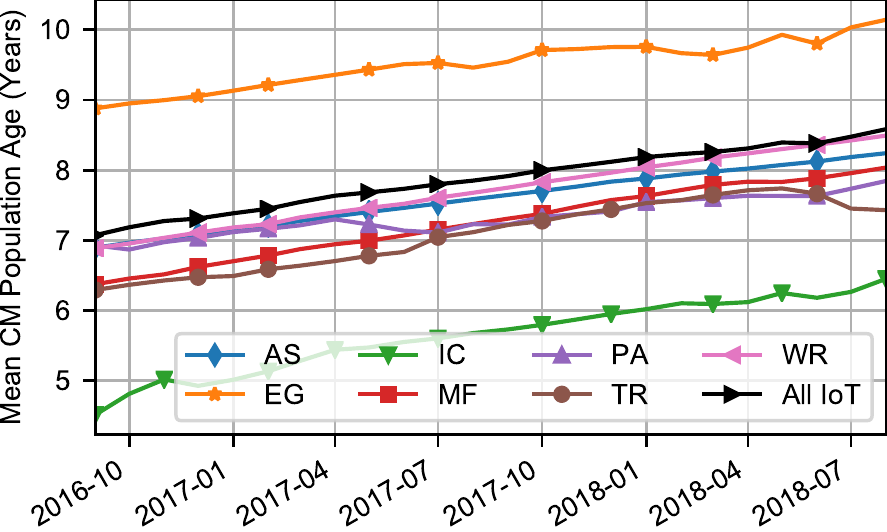}
\caption{Mean CM population age (in years) based on the CM model release year}
\label{fig:device_population_age}
\end{figure}

Through leveraging the additional device features dataset, we can analyze the feature and CM age evolution of the \iotname device population. Figure \ref{fig:device_population_age} shows the evolution of the mean CM age by industry, with age defined as the time since the release year of the CM model\footnote{We assume that CM models are released on Jan. 1st. In other words, we overestimate the actual age, but this does not preclude tracking temporal dynamics and comparing industries.} (and not the manufacturing year of the CM). We observe that the mean CM age was over 8.5 years, as of August 2018. The {\it Electricity and gas} industry has the oldest CM population, with a mean age of more than ten years. Overall, the increase in mean population age for all industries illustrates the slow pace of new CM model deployment. We further analyzed the population of CMs deployed after September 2016 and found that the mean age in August 2018 was about seven years. 

\begin{figure}
\centering
\includegraphics[width=\picwidth]{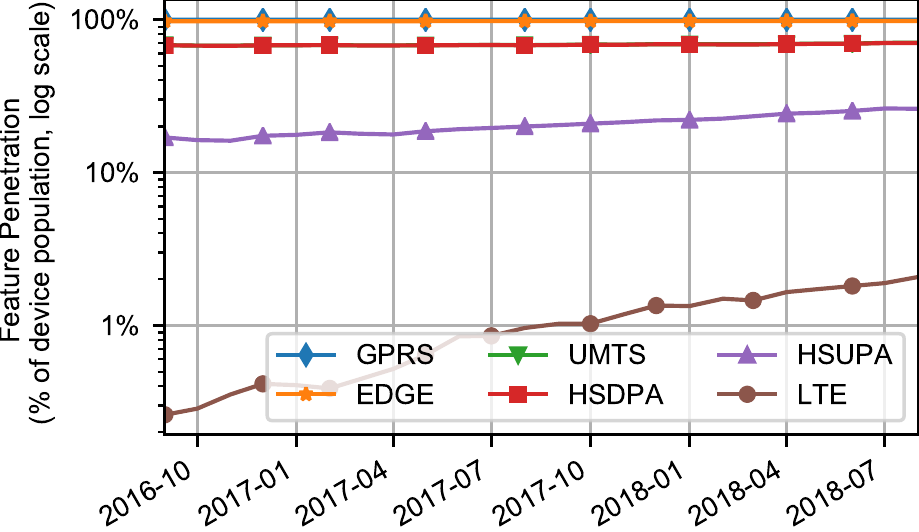}
\caption{Penetration of \iotname device features}
\label{fig:feature_diffusion}
\end{figure}

In terms of connectivity features, Figure \ref{fig:feature_diffusion} presents the penetration of 3GPP connectivity technologies among the \iotname device population. First, we observe the low penetration (and growth rate) of LTE of about 2\% as of August 2018. This observation is in line with the significant age of the \iotname device CM population and contrasts with LTE penetration of 41\% among non-\iotname devices in Europe in 2017 \cite{GSMAssociation2018}. {\color{black} Additionally, the growth rate of LTE among Finnish smartphones was seven times as large as \iotname devices over a comparable yearly time-frame (specifically when re-basing both series to start at 0.1\% penetration) \cite{vesselkov2014}.}

We further find that although the penetration of LTE is growing in all industries, only in the {\it Transportation} industry has penetration exceeded 10\%. Furthermore, we observe a significant difference in the penetration of HSDPA and HSUPA technologies of 70\% and 26\% respectively. This is surprising given the prevalence of uplink traffic in \iotname which suggests a stronger need for fast uplink technologies rather than downlink. Finally, we find the share of 2G only (GPRS and EDGE) devices is still about 30\%. Therefore, discontinuing 2G service (for spectrum reuse purposes) would indeed affect a significant fraction of \iotname devices thus posing a problem for network operators.

\begin{figure}
\centering
\includegraphics[width=\picwidth]{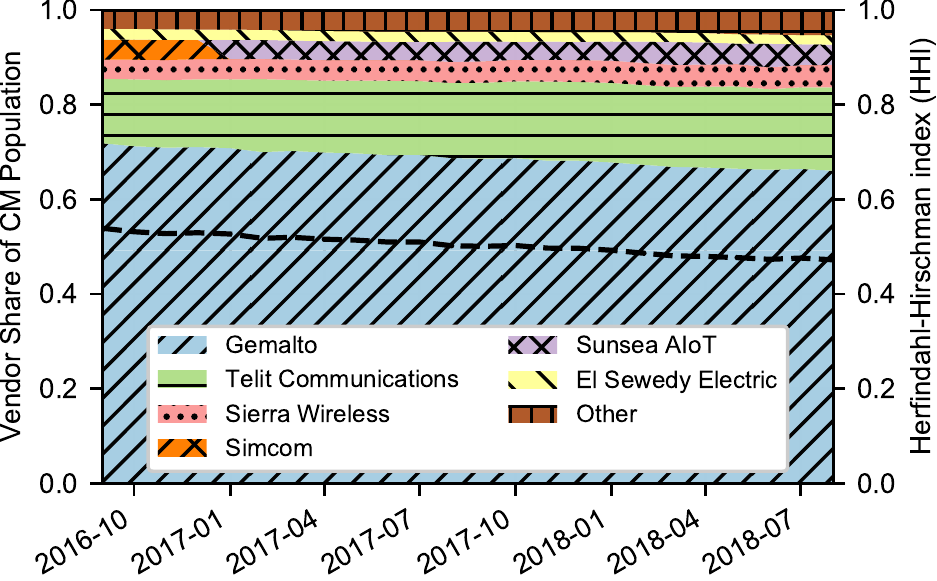}
\caption{{\color{black}Vendor share of CM population and the Herfindahl-Hirschman index (HHI)}}
\label{fig:vendor_market_share}
\end{figure}

{\color{black}
We also examine the prevalence of NB-IoT/LTE-M capable devices in the population by using a publicly available list of such devices from GSMA\footnote{https://www.gsma.com/iot/mobile-iot-modules/}. However we find that these devices represent less than 0.05\% of all devices and thus are too small of a sample for reliable analysis. Furthermore the devices currently using the network are likely primarily testing devices.

We examine the shares of CM vendors in the device population over the observation period\footnote{We gather CM vendor information including public mergers, acquisitions, and divestment data (including dates) for all CM vendors that appear in the device population.}. Figure \ref{fig:vendor_market_share} illustrates the CM vendor shares over time and the Herfindahl-Hirschman index (HHI), a measure of market concentration. The figure indicates quite high concentration with only a small decrease in concentration as quantified by HHI (from 0.53 to 0.47) over the observation period. Though, we note that if the customer company with the most IoT devices is removed (to assess sensitivity), the HHI for August 2018 drops to 0.30, thus illustrating the potential for volatility given large customers.}

\section{Modeling and Prediction Results}\label{sec:adv_results}

\subsection{Temporal traffic spectrum and clustering analysis}
To obtain additional temporal traffic insights, we perform several types of temporal analysis on different time scales. Specifically, our approach is to study the short timescale (hours, weeks) temporal patterns of three different months roughly evenly spaced over the two years: September 2016, August 2017, and August 2018. We always present the results from August 2018 and only present and note the results from the earlier months if substantially different.

\begin{figure}
\centering
\includegraphics[width=\picwidth]{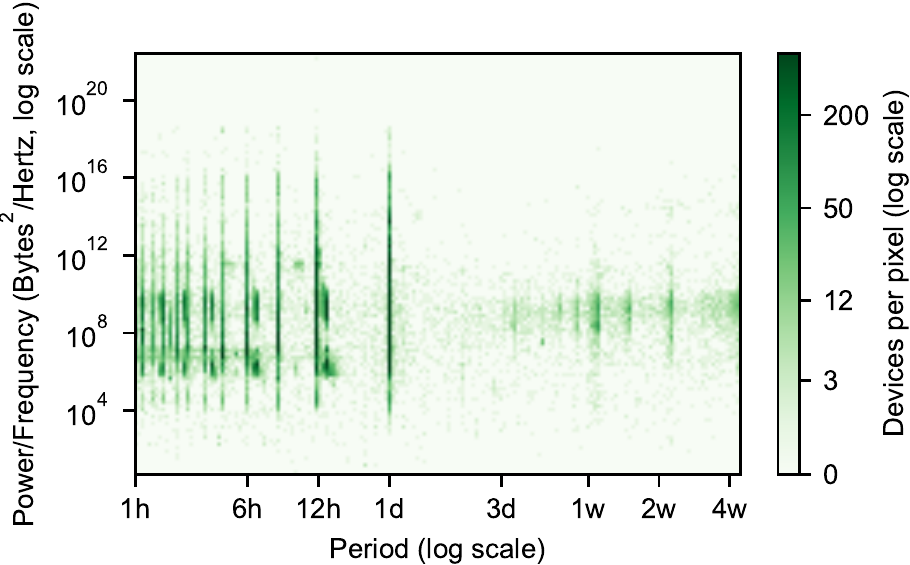}
\caption{Density of spectrum peaks vs periods of devices for total traffic for August 2018}
\label{fig:periodogram_data_in_peaks}
\end{figure}

First, we perform spectral analysis on a one-month traffic volume series of each device for uplink and downlink traffic. The spectral density of each series is estimated as the squared modulus of the discrete Fourier transform, in other words the periodogram. Then the peak power and corresponding period are extracted from each periodogram under the assumption that most \iotname devices will have a dominant timer-driven peak. Figure \ref{fig:periodogram_data_in_peaks} illustrates the density of these (peak power, period) pairs for downlink traffic. The plot for uplink is almost identical. We find large fractions of devices have peaks at 24, 12, and 6 hour periods including devices with large and small peak traffic volumes (power). However, we also find other periods such as \texttildelow13 hours, though this case is specific to only two large companies with similar device models. The reason for the use of a 13 hour period in these companies is unknown. We also note that some devices have peaks at one week thus reinforcing the patterns from Figure \ref{fig:data_per_day}, however these devices tend to have small peak traffic volumes.

For a more granular temporal analysis, we perform temporal clustering on the averaged (over the month) and normalized 24-hour total traffic volume series of each device. The normalization is performed for each device over the 24-hour series such that the value for any given hour is the fraction of that device's total daily traffic in that hour. This normalization is required due to the order of magnitude differences in traffic volumes between some devices. Each series is then transformed by a discrete wavelet transform (DWT) with a Daubechies-1 wavelet and a decomposition level of three.

The DWT coefficients are then clustered via bisecting k-means with the number of clusters chosen by the silhouette score. We use bisecting k-means because of the $O(n)$ run-time and ease of computational distribution. Comparatively, other approaches such as hierarchical clustering with ward linkage have a run-time of $O(n^{2})$. Though for robustness, we also cluster a random sample of 2000 devices via hierarchical clustering with ward linkage and with the number of clusters chosen by the Davies-Bouldin score. This is the same clustering setup as in \cite{shafiq2013}. We find the same number of clusters as the full device clustering and virtually the same cluster centroids.

\begin{figure}
\centering
\includegraphics[width=\picwidth]{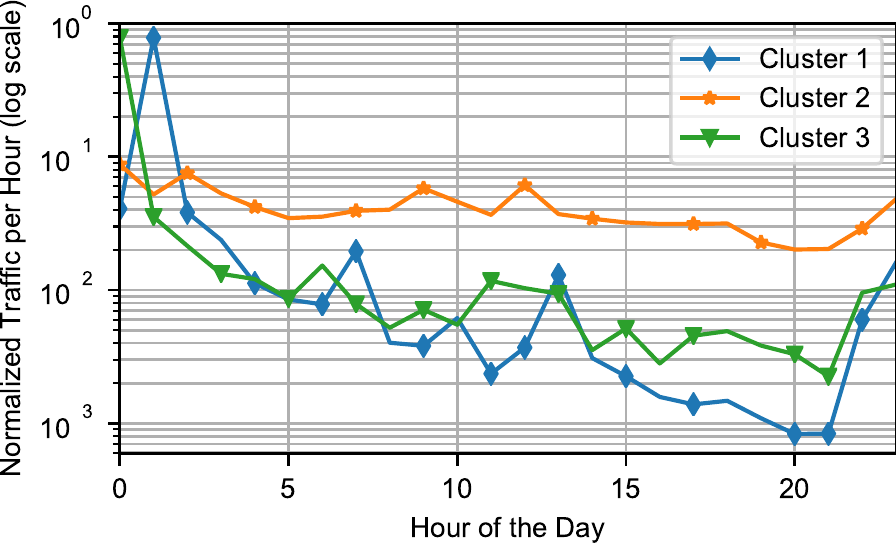}
\caption{Cluster centroids of the three temporal clusters from August 2018}
\label{fig:cluster_centroids}
\end{figure}

Regarding clustering results, we find that the optimal number of clusters is three. The clusters denoted 1, 2, and 3 encompass 25\%, 41\%, and 34\% of devices respectively. The cluster centroids (in terms of time series rather than DWT coefficients) of the three clusters are illustrated in Figure \ref{fig:cluster_centroids}. We find that clusters 1 and 3 have significant peaks at 0:00 and 2:00 respectively with over 80\% of their traffic within that peak hour, while cluster 2 shows much steadier and flatter traffic throughout the day.

To better understand these patterns we look at the composition of the clusters by company ID and industries. Interestingly, 88\% of cluster 1 devices belong to a single large company; thus this cluster is company-specific and not necessarily a general \iotname temporal pattern. Though \cite{shafiq2013} also found an outlier cluster with a peak at 02:00 that they attributed mainly to fleet management applications. For clusters 2 and 3, no single company represents more than 31\% of devices and no single industry more than 50\% of devices. The main industries for cluster 2 are {\it Wholesale and retail trade} (40\%), {\it Electricity and gas} (22\%), and {\it Information and communication} (14\%), while the main industries for cluster 3 are {\it Wholesale and retail trade} (51\%), {\it Electricity and gas} (30\%), {\it Administrative and support services} (11\%). This overlap in industries highlights diversity in use cases even within narrow industries such as {\it Electricity and gas}.

\begin{figure}
\centering
\includegraphics[width=\picwidth]{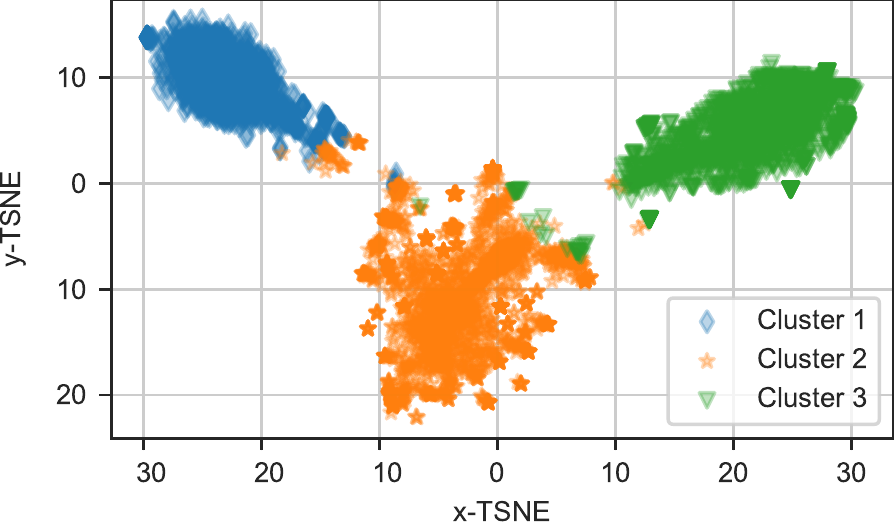}
\caption{T-SNE of sample of 4400 devices from the three temporal clusters from August 2018}
\label{fig:tsne_results}
\end{figure}

For an illustration of cluster separation, we plot the t-distributed stochastic neighbor embedding (t-SNE) of a random sample\footnote{t-SNE has a run-time complexity of $O(n^{2})$ and thus does not scale to large data.} of 4400 devices in Figure \ref{fig:tsne_results} with perplexity chosen as in \cite{cao2017}. The clusters appear to be well separated with only minimal overlap, especially the single-company dominated cluster 1, thus reinforcing the clustering results.

In terms of longer scale temporal phenomena, we did not find large differences in either temporal analysis method between the examined months. This suggests that \iotname phenomena change slowly; such behavior reinforces the previously identified slow change in, for example, device feature penetration.

\begin{figure}
\centering
\includegraphics[width=\picwidth]{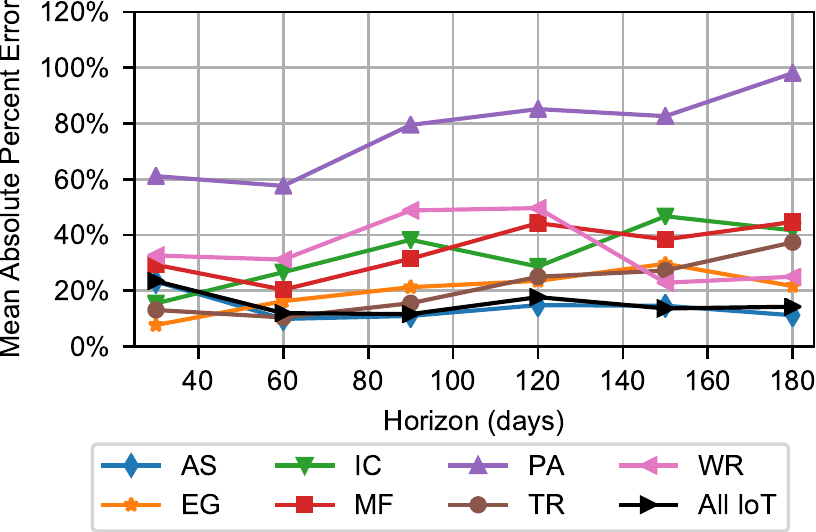}
\caption{{\color{black} Estimated MAPE of forecasts over varying time horizons (30 to 180 days) for all \iotname and industry-specific models}}
\label{fig:forecasting_performance}
\end{figure}

{\color{black}
\subsection{Temporal traffic forecasting}
Finally, we examine the possibility of \iotname traffic forecasting (a common network operator goal). Specifically, we evaluate the potential for medium-term forecasting of daily \iotname traffic through a state of the art forecasting model. This medium-term (from 30 to 365 day horizons) forecasting is the most practical given the length of our observation period (about two years) and granularity (hourly).

Namely, we use the Prophet time series model \cite{taylor2018} which is a decomposable additive regression model. Equation \ref{eq:prophet_model} details the high-level model formulation consisting of piece-wise (linear or logistic growth) trends $g(t)$, seasonality (weekly or yearly) $s(t)$, and holiday $h(t)$ components plus an error term $\epsilon_{t}$. The model fitting is flexible and automatically selects the appropriate trend change points and components. Specifically, the model fitting is performed through the probabilistic programming language Stan \cite{carpenter2017} via maximum a posteriori parameter estimation. The holiday component of the model uses the national holidays of Finland. The main justification for using Prophet over alternative models such as autoregressive integrated moving average (ARIMA) is that Prophet was designed for web event modeling and thus natively supports common network traffic properties such as the aforementioned piecewise trends, seasonality, and holidays.

\begin{equation}\label{eq:prophet_model}
y(t)=g(t)+s(t)+h(t)+\epsilon_{t}
\end{equation}

Firstly, for reference Figure \ref{fig:forecasting_reference} illustrates the daily traffic per device series for all \iotname devices (hereafter All-\iotname), the model fitted on that series for the entire observation period, and a one year forecast. The model illustrates two distinct trends with a visible change point at Oct. 2017 and a weekly seasonality that aligns with a weekday/weekend dichotomy. Visually, the model appears to provide a simple though reasonable fit of the series.

\begin{figure}
\centering
\includegraphics[width=\picwidth]{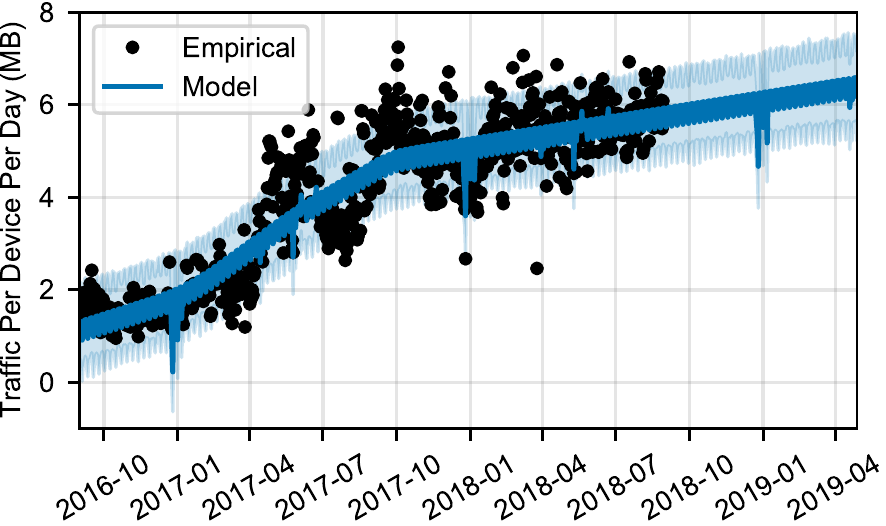}
\caption{{\color{black} All-\iotname daily traffic per device and fitted model including one year forecast}}
\label{fig:forecasting_reference}
\end{figure}

Next, to estimate the accuracy of forecasting we use a rolling window validation method known as simulated historical forecasts (SHF) \cite{taylor2018}. In the SHF method forecasts are made for rolling historical horizons given a fixed training window size, a variable horizon window size, and a fixed period for shifting these windows within the full historical window. We use a training window size of 365 days, horizon window sizes from 30 to 180 days, and a period of 90 days. This allows the estimation of forecasting accuracy for time horizons ranging from 30 to 180 days. This estimation is performed for the all-\iotname daily traffic per device series and each daily industry series individually. For easy interpretation, accuracy is quantified by the mean absolute percent error (MAPE).

In terms of results, Figure \ref{fig:forecasting_performance} details the estimated MAPE for forecasts over different time horizons for both the all-\iotname and industry-specific models. Interestingly, accuracy for the all-\iotname model is respectable with an error between 10-20\%; whereas for the different industries the model accuracies vary more substantially. The low accuracy for the {\it Professional activities} industry is due to the high series variance and lack of a clear trend (see again Figure \ref{fig:data_per_imsi_industry_two_years}) suggesting that the industry may be too diverse to be a useful as an aggregation. Generally, as expected, accuracy decreases with longer time horizons. The overall moderate accuracy implies that network operators could use such models at least for general high-level planning of \iotname usage in the medium-term.

Additionally, forecasting of individual customer company traffic over time may be useful especially for large companies. To assess the viability of such forecasting, we perform the same procedure as previously but for the ten largest companies by number of devices. These companies are from six different industries and thus relatively diverse. We find the mean MAPEs (over the 30-180 day horizons) for the companies range between 2\% to 55\%; therefore illustrating similar diversity as on the industry level. This also reinforces that the traffic variation that impinges forecasting is both intra-industry and intra-company.

For research purposes we release publicly the all-\iotname and industry specific models as serialized Python pickle files\footnote{The models can be found at (link omitted for double blind review purposes).}. These files can be imported into the Prophet library to allow forecasting and interrogation of the parameters. 
}

{\color{black}
\subsection{Mobility modeling and analysis}
Next, we perform mobility modeling to capture the typical patterns \iotname devices follow across cells. To do so we utilize a finite mixture model of Markov models fit through a library known as Clickclust \cite{melnykov2016}. Each Markov model represents the transitions between a fixed set of categories (specifically cells).

More specifically, Clickclust \cite{melnykov2016} estimates a finite mixture model of Markov models for a set of categorical sequences where the probability distribution of the finite mixture model is 
\[f(\gamma |\vartheta ) = \sum\limits_{k = 1}^K {{\alpha _k}{f_k} (\gamma |{\vartheta _k})}\]

where $K$ is the number of component distributions $f_k(\cdot |{\vartheta _k})$ with parameter vectors (and mixing proportions $\alpha_1,\ldots,\alpha_{K-1}$) subject to restrictions $\alpha_{k}>0$ and $\sum\limits_{k = 1}^K {\alpha_{k}}=1$. Each component distribution is a first order Markov model representing a cluster of similar cell sequences. The number of component distributions $K$ is selected through agglomerative clustering to minimize the Bayesian information criterion (BIC) via a two-stage iterative procedure with an expectation-maximization (EM) algorithm.

Due to computational complexity issues, the assumption of diurnal patterns, and the desire to model mobile (rather than mostly stationary) devices, we perform some initial processing and filtering. Specifically, we only model the cell sequences of devices from September 30 to 31, 2018. We also remove runs of the same cell in the cell sequences and thus focus only on the notion of mobility. We then select a random sample of 2000 devices with a cell sequence length of at least five (i.e., sent or received in at least five hour-cell combinations) and a cell sequence with between three and 50 distinct cells. Finally, we normalize each cell sequence by encoding the most frequent cell as 0, the next most frequent as 1, and so on.

In terms of results, a simple application of the EM algorithm suggests an optimal mixture model with one component, as shown in Table \ref{tab:mobility_modeling_results} (where BIC is minimized at $K=1$). This result suggests potential model over-parameterization and order underestimation, which is possible with a large Markov state space (our state space is 50 due to the distinct cell limit from our filtering). Fortunately, Clickclust contains a forward state selection (FSS) algorithm, allowing for the aggregation of Markov states into equivalence blocks considering their transition probabilities. The application of the FSS procedure estimates an optimal mixture model with three components with a BIC of 66014.44. Table \ref{tab:mobility_modeling_results} provides a performance summary for both the EM and FSS procedure\footnote{For reference, the EM and FSS procedures use the following parameters iter = 3, eps = 1e-8, r = 50, min.gamma = 1e-2, and min.beta = 1e-2. The computation time for FSS for K=4 was approximately ten hours on eight Intel Xeon X5650 CPUs with 80GB total RAM.} including the number of equivalence blocks $d$ for FSS.

\begin{table*}[ht]
\small
\caption{{\color{black}Clickclust mixture model fitting BIC scores for different numbers of Markov models (in parentheses is the number of equivalence classes in FSS)}}
\label{tab:mobility_modeling_results}
\begin{tabular}{lllll}
\hline
\textbf{Method} & \textbf{K = 1} & \textbf{K = 2} & \textbf{K = 3} & \textbf{K = 4} \\
\hline
EM & \textbf{88203.11} & 104810.3 & 121983 & 139020.2 \\
\hline
FSS & 66961.61 (9) & 66236.89 (8) & \textbf{66014.44} (7) & 66177.28 (6) \\
\hline
\end{tabular}
\end{table*}

The optimal solution provided by the FSS procedure has the mixing proportions $\alpha _1,\alpha _2,\alpha _3$ of 0.67, 0.04, 0.29, indicating an unbalanced weight distribution for the components. The FSS procedure aggregated the 50 distinct cells into seven equivalence blocks, as shown in Table \ref{tab:forward_map}. We observe that the number of distinct cells per block increases when including less visited cells, indicating that transition probabilities among less visited cells are more similar than among more visited cells.

\begin{table*}[ht]
\small
\caption{{\color{black}Mapping of distinct cells to equivalence blocks in FSS}}
\label{tab:forward_map}
\begin{tabular}{l|l}
\hline
\textbf{Cell Order} & \textbf{Equivalence Blocks} \\
\hline
1st-25th most visited & 4 5 6 7 3 3 3 3 2 2 2 2 2 2 2 2 2 2 1 1 1 1 1 1 1\\
\hline
25th-50th most visited & 1 1 1 1 1 1 1 1 1 1 1 1 1 1 1 1 1 1 1 1 1 1 1 1 1\\
\hline
\end{tabular}
\end{table*}

Figure \ref{fig:transition_matrices} illustrates the transition probabilities between equivalence blocks for each component of the mixture model. Component 1, which has the highest weight, mainly models the bidirectional transitions between the two most visited cells (equivalence blocks 4 and 5, as shown in Table \ref{tab:forward_map}). Component 2, which has the smallest weight, primarily models the transitions between the least visited cells (blocks 1 and 2) and other blocks. Finally, component 3 models transitions between multiple other blocks excluding block 1. The high transition probabilities between the top cells may partly be the result, as discussed earlier, of devices at cell edges that jump between cells depending on signal strength fluctuations or cell load balancing. Remember also that sequences with less than three unique cells were removed from the mobility modeling.

\begin{figure*}[ht]
\centering
\includegraphics[width=\picwidth]{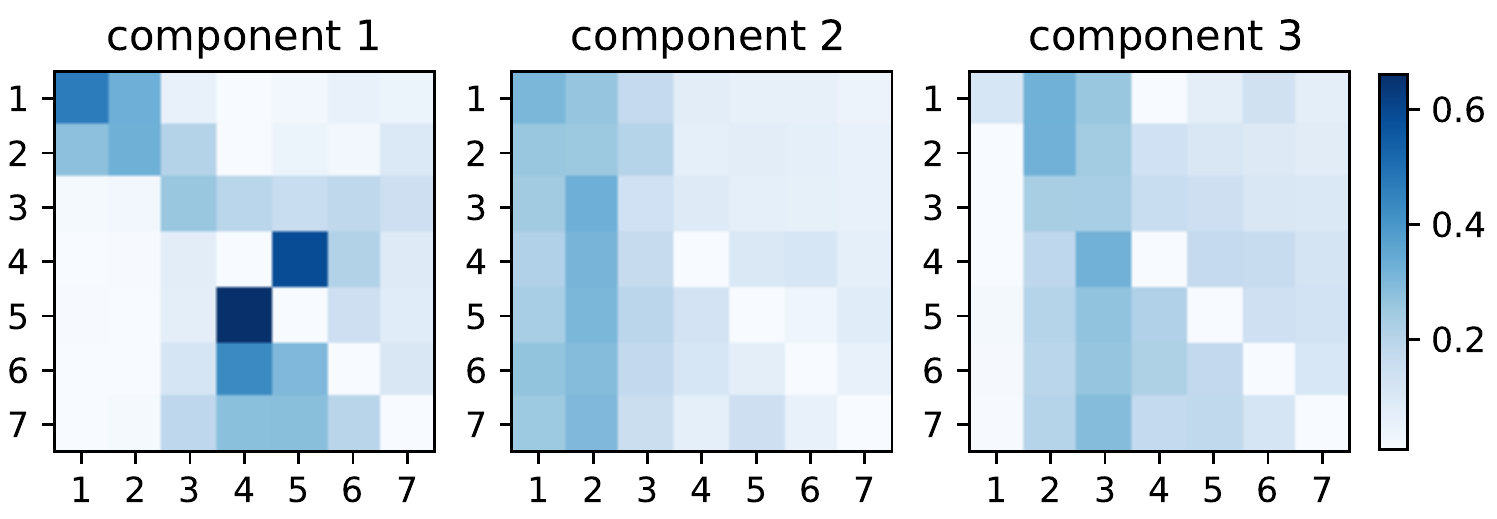}
\caption{{\color{black} Transition matrices between equivalence blocks for each component of the FSS model}}
\label{fig:transition_matrices}
\end{figure*}

For research purposes we release publicly the 3-cluster FSS mixture model and script\footnote{The models can be found at (link omitted for double blind review purposes).} (as serialized R objects and files) thus allowing for model interrogation and the synthetic generation of cell sequences. Transition matrices can also be found within the script.

We also perform an alternative mobility analysis using an information theory framework. Specifically, we quantify for all devices the potential for next cell prediction and optimization by estimating the information theoretical upper bound of predictability (hereafter $\Pi^{max}$) similar to \cite{song2010}. $\Pi^{max}$ denotes the theoretical maximum percentage of cell visits that could be predicted given the entropy of the cell sequence. Though in contrast to \cite{song2010} we only estimate the predictability of cell attachments with data transfer (as this is our definition of cell visit) rather than of all cell attachments (which would infer the full mobility as in \cite{song2010}). This formulation avoids a common missing data problem from prior work in that cell attachments without data transfer do not generate CDRs or DDRs and are therefore often absent from mobile network datasets.

The entropy of a sequence is estimated through a Lempel-Ziv compression based estimator detailed in Equation \ref{entropy_estimator} where $n$ is the length of the sequence and $\Lambda_{i}$ is the length of the longest subsequence starting from $i$ and not seen earlier from 1 to $i-1$. This estimator quickly converges to the true entropy rate as $n\rightarrow \infty$. This entropy is then used to numerically solve for $\Pi^{max}$ through Equation \ref{upper_bound} (which is derived from Fano's inequality \cite{song2010}) where $N$ is the number of distinct symbols (i.e., cells) in the sequence.

\begin{equation}\label{entropy_estimator}
H_{rate}=\left(\frac{1}{n} \displaystyle\sum_{i=1}^{n} \frac{\Lambda_{i}}{\log_2{n}} \right)^{-1}
\end{equation}

\begin{equation}\label{upper_bound}
\begin{split}
& H_{rate}=-\Pi^{max}\log_2{(\Pi^{max})} -
 (1-\Pi^{max})\log_2{(1-\Pi^{max})} + \\
& (1-\Pi^{max})\log_2{(N-1)}
\end{split}
\end{equation}

We examine the single cell sequence covering from March to August 2018 (last six months of the observation period). As before we remove runs of the same cell in the cell sequence. We also limit the analysis to devices with a cell sequence length of at least 20. The longer observation period and length limit are necessary because the entropy estimator requires a reasonable length for accurate estimation\footnote{Specifically, the variance and bias decrease proportionally as $1/n$ and $1/log_{2}{(n)}$ respectively. The threshold of 20 is somewhat arbitrary, unfortunately determining the sequence length required for a specific entropy estimation accuracy is non-trivial and current methods (i.e \cite{back2019}) assume i.i.d. and Zipfian symbol probabilities.}. This limitation removes lower activity, mostly stationary devices (about 39\% of devices).

Figure \ref{fig:entropy_precitability} illustrates the ECDF of $\Pi^{max}$ for devices by industry for the period of March to August 2018. The results illustrate that a non-trivial fraction of \iotname devices have unpredictable cell mobility dynamics; especially in the {\it Transportation} and {\it Information and communication} industries. This suggests that network operators could first focus their predictive optimization efforts on industries with high predictability to gain quick wins.

\begin{figure}
\centering
\includegraphics[width=\picwidth]{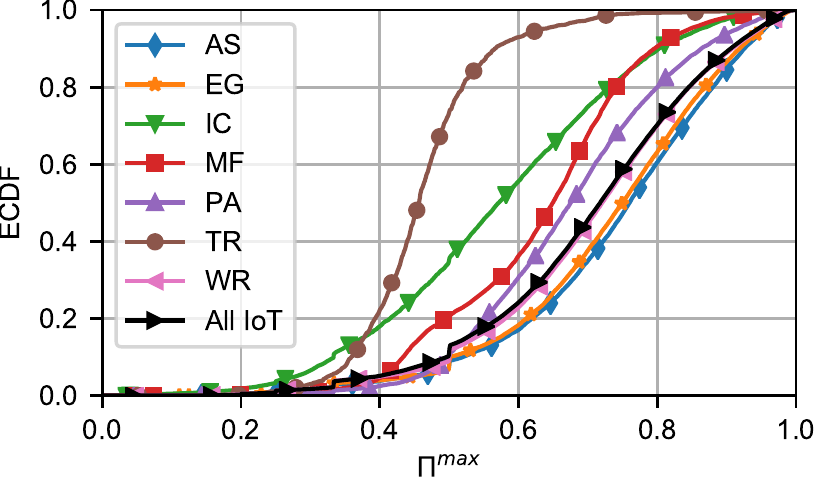}
\caption{{\color{black} ECDF of $\Pi^{max}$ (upper bound of predictability) for devices by industry for period of March to August 2018.}}

\label{fig:entropy_precitability}
\end{figure}
}

\section{Limitations}\label{sec:limitations}
We next discuss two major study limitations. First, the study is not fully representative since the dataset is from only a single MNO in a single country and includes only \iotname and not consumer IoT. However, as previously mentioned, we hope that similar studies from other countries and MNOs can help to build up a wide-ranging and practical understanding of IoT dynamics. Second, the time resolution of one hour means that the study might have missed more granular phenomenon on the minute and second timescales. However, we note that potentially more intrusive data collection methods would be required for those timescales, thus potentially hindering collection. Therefore, we leave such granular work for future studies.

\section{Conclusions}\label{sec:conclusions}
Overall, this work presented an analysis of cellular \iotname traffic and mobility patterns over several different timescales for a major Finnish MNO. The analysis includes trends over a two-year span thus allowing a view of the evolution of \iotname. Moreover, trends were broken down by industry, and the penetration of device features in the \iotname device population was analyzed. Finally, the analysis evaluated \iotname traffic forecasting and mobility modeling. Overall the analysis provided a diverse set of results of which we highlight a few.

For example, we found that \iotname traffic per device tripled over the last two years; however, the mean age of CM models in the device population also increased significantly to over eight years. Furthermore, the penetration of LTE-enabled \iotname devices is very low (2\%) and growing very slowly. Also we found significant variation between devices of different industries with orders of magnitude differences in traffic volume and mobility. {\color{black} Furthermore, we illustrated that total daily \iotname traffic can be accurately forecast (\textasciitilde15\% error) over a medium-term (30 to 180 day) horizon. Finally, we presented that a non-trivial fraction of \iotname devices have inherent unpredictability in terms of their mobility.}

{\color{black}
The results have implications for mobile ecosystem players. We note the following example implications:
\begin{itemize}

\item MNOs should be cautious in discontinuing 2G service (for spectrum refarming purposes) since a large fraction of \iotname devices likely still use GPRS and EDGE and the \iotname device life cycle is lengthy.

\item Network managers should consider \iotname spatio-temporal traffic patterns when defining mesoscale (on order of hours or days) IoT network configurations and optimizations (such as the powering down of certain BSs for energy saving purposes) or when developing ML models that perform such optimizations. For example, given the very high spatial traffic concentration and thus inter-base station variability, the use of cell level or multi-level (rather than global) ML models is likely to be important.

\item Network planners should consider the specific requirements of those industries targeted by business development plans given the significant inter-industry traffic and mobility differences. An example of relevant planning could be the optimal placement of resources for edge computing. Specifically, in the administrative and support industry, uplink capacity could be saved by enabling the edge processing of video streams from CCTV systems with no built-in ML object detection. 

\item Industry customers, MNOs, and \iotname device manufacturers should collaborate with multiple CM vendors to avoid the risk of over-reliance on a single vendor (given the potential for high CM market concentration).
\end{itemize}
}

\section{Acknowledgments}
This work has been supported by the Digital Disruption of Industry project funded by the Strategic Research Council of Finland (Grant No. 292889), the 5GEAR project funded by the Academy of Finland (Decision No. 319669), and the FIT project funded by the Academy of Finland (Decision No. 325570), and the 5G-VIIMA project funded by Business Finland (Grant no. 6430/31/2018). We also acknowledge the computational resources provided by the Aalto Science-IT project.

\bibliographystyle{elsarticle-num}
\bibliography{refs.bib}
\end{document}